\documentclass[
aps,
 ,twocolumn
]{revtex4}
\usepackage{graphicx}

\begin{document}

\title{Reply to comment on ``Information Flow of quantum states interacting with closed timelike curves"}

\author{T.C.Ralph and C.R.Myers}\affiliation{
School of Mathematics and Physics, University
of Queensland, St Lucia, Queensland 4072, Australia}


%
\begin{abstract}

We respond to the comment by K{\l}obus, {\it et al} by emphasizing that the equivalent circuit, once constructed, obeys the standard rules of quantum mechanics - hence there is no ambiguity in how to choose initial states in our model. We discuss the distinction between correlated ensembles produced non-locally via measurements on entangled states and those produced via local preparation.

\end{abstract}


\maketitle


\vspace{10 mm}

In our recent paper, Ref.\cite{RAL10}, we introduced the equivalent circuit approach to solving quantum evolution in the presence of closed timelike curves (CTCs). The equivalent circuit represents the dynamics of the system as viewed from the perspective of the qubit traversing the CTC. The equivalent circuit is constructed by mapping: a single pure state input, $|\phi \rangle$, in the CTC system, to $n$ identical copies, $|\phi \rangle^{\otimes n}$, in the equivalent circuit \cite{Note}; and a single unitary interaction between past and present incarnations of the qubit in the CTC system to $n$ identical copies of the interaction in the equivalent circuit (see Fig.1 Ref.\cite{RAL10}). Formally we allow $n \to \infty$. The copies represent the looping back in time that characterize the CTC. In order to retrieve standard quantum mechanics in the absence of a CTC, it is necessary to assume that only one of the $n$ input modes is eventually detected - corresponding to the single output mode of the CTC system. Having constructed it via the mapping described above, the equivalent circuit is solved by applying the {\it standard rules of quantum mechanics}.

Using this construction we were able to independently derive the density operator consistency requirements introduced by Deutsch \cite{DEU91} for modelling quantum systems interacting with CTCs. The advantage of this new derivation was that we were able to resolve two ambiguities in the Deutsch formulation: (i) the question of how to treat classically correlated input states; and (ii) the question of how to choose the correct solution in situations where multiple solutions appear.
In their comment on our paper, K{\l}obus, Grudka, and W\'{o}jcik \cite{KLO11} contend that the first of these ambiguities is not resolved by the equivalent circuit. 

In the Deutsch formulation it is the reduced density operator which is matched across the CTC boundary. This requires the tracing out of all modes other than those being matched, in order to decide the consistent solution. It was pointed out by Deutsch in his original paper that this procedure can lead to the de-correlation of entanglement. However he did not discuss what happens to classically correlated systems. Most authors have assumed that classical correlations are preserved (eg: Refs \cite{BAC04, BRU09}), however others have argued they would be destroyed \cite{BEN09}. Mathematically, this comes down to deciding whether the trace should be taken on a shot-by-shot basis, or on the entire ensemble. 

The issue is resolved in the equivalent circuit formalism by simply applying the standard rules of quantum mechanics.
K{\l}obus, {\it et al} argue that there is an ambiguity about how to represent classically correlated initial states in the equivalent circuit formalism and show that destruction of classical correlations can occur when a different form for the initial states is used. In Ref.\cite{RAL10}, considerable time is spent deriving the correct, unique form for the initial state, in various different situations. Strangely, no attempt is made in Ref.\cite{KLO11} to counter, or even acknowledge this derivation. Instead K{\l}obus, {\it et al} simply claim that an arbitrary choice can be made. 

K{\l}obus, {\it et al} agree that a pure ensemble of states in the $|0 \rangle|0 \rangle$ state should be represented
by 
\begin{equation}
\rho_0 = (|0 \rangle|0 \rangle \langle 0|\langle  0|)^{\otimes n}
\end{equation}
in the equivalent circuit and similarly a pure ensemble in the $|1 \rangle|1 \rangle$ state should be represented by
\begin{equation}
\rho_1 = (|1 \rangle|1 \rangle \langle 1|\langle 1|)^{\otimes n}
\end{equation}
in the equivalent circuit. Given this, standard quantum mechanics tells us that an equal mixture of these states is given by:  
\begin{eqnarray}
\rho & =& \frac{1}{2}(\rho_0+\rho_1) \nonumber\\
&=& \frac{1}{2}((|0 \rangle|0 \rangle \langle 0| \langle 0|)^{\otimes n} + (|1 \rangle|1 \rangle \langle 1|\langle 1|)^{\otimes n})
\end{eqnarray}
as used in
our paper. There is no ambiguity. Any other choice (in particular the choice in Ref.\cite{KLO11}) is inconsistent with standard quantum mechanics. Imposing non-standard conventions onto a standard circuit inevitably breaks the internal
self-consistency of quantum mechanics. For example, the standard formalism for describing classically mixed states allows a consistent description of an experiment in terms of mixed states, or sub-ensembles of pure states. Thus the preparer of the states - who knows the shot by shot input states - will predict outcomes that are consistent with those predicted by the measurer of the states - who knows only the statistics. This is not the case for the non-standard convention used in Ref.\cite{KLO11}.

An interesting example raised in Ref.\cite{KLO11} is that of their Eq.5. We might characterize this example as the exception that proves the rule. They consider producing a ``classically correlated" state by making measurements on one member of a maximally entangled Bell pair before it interacts with a CTC. We will refer to this as correlation via entanglement, as opposed to correlation via preparation. Clearly these two situations are physically distinct - in the former the correlation is produced non-locally and the experimenter has no control over what specific states are produced shot-by-shot, in the latter the correlation is produced locally by specific choices of the experimenter. Never-the-less, in the absence of CTCs, quantum mechanics treats these cases equivalently. K{\l}obus, {\it et al} demonstrate that our formalism makes a distinction between these two cases, de-correlating the former but not the latter. The implication is that this example implies an inconsistency in our formalism.

As previously discussed, it is not controversial that entanglement can be de-correlated by the CTC. However, what if, as K{\l}obus, {\it et al} suggest, the entanglement is ``collapsed" by a measurement before the interaction with the CTC. Does this now constitute classical correlations that should not be de-correlated? After some thought, it is obvious that such a situation would be inconsistent with special relativity. Consider the situation in which the correlating measurement was space-like separated from the CTC. Inertial observers in different reference frames could observe a different ordering of the measurement event versus the entry of the other qubit to the CTC, and as a result, would predict distinct and contradictory outcomes. 

Thus self-consistency suggests that states that are correlated via entanglement should be de-correlated by the CTC. In some sense this must be put in by hand to the Deutsch formalism. In contrast, as K{\l}obus, {\it et al} show, it emerges naturally from the equivalent circuit formalism. This is perhaps not so surprising - the equivalent circuit is a physical circuit obeying the rules of standard quantum mechanics and so is guaranteed to be consistent with special relativity.

In summary, it appears that K{\l}obus, {\it et al}  may have misunderstood a key point of the equivalent circuit formalism. They state ``...at the moment it is not known what the dynamics of states interacting with closed timelike curves depends on." The point of our formalism is to map these ``unknown dynamics" onto an equivalent circuit for which the dynamics {\it are} known. Their counter example - illustrating the difference between correlations via entanglement and correlations via preparation - actually illustrates the self-consistency of the formalism.


\begin{thebibliography}{99}

\bibitem{RAL10} T.C.Ralph and C.R.Myers, Phys Rev A {\bf 82}, 062330 (2010).

\bibitem{Note} If the input state is entangled then we assume it is the pure, multi-mode entangled state which is copied (see Ref.\cite{RAL10} for details). It is a standard, though perhaps not universally accepted, assumption that such a purification can always be performed.

\bibitem{DEU91} D.Deutsch, Phys.Rev.D {\bf 44}, 3197 (1991).

\bibitem{KLO11} W.K.K{\l}obus, A.Grudka, and A.W\'{o}jcik, Phys. Rev. A {\bf 84}, 056301 (2011).

\bibitem{BAC04} D.Bacon, Phys.Rev.A, {\bf 70} 032309 (2004).


\bibitem{BRU09} T.A.Brun, J.Harrington, and M.M.Wilde, Phys. Rev. Lett. {\bf 102}, 210402 (2009).

\bibitem{BEN09} C.H.Bennett, D.Leung, G.Smith and J.A.Smolin, Phys Rev Lett {\bf 103}, 170502 (2009).


\end{thebibliography}
\end{document}